\documentstyle[]{l-aa}

\begin{document}
\input{psfig.sty}

\thesaurus{10.07.2; 11.05.1; 11.09.1 NGC 1399, NGC 1374, NGC 1379, NGC
1387, NGC 1427; 11.03.04 Fornax}

\title{Globular cluster systems of early--type galaxies in
Fornax
\thanks{Based on data collected at the Las Campanas Observatory,
	 Chile, run by the Carnegie Institutions}
}

\author {M.~Kissler--Patig \inst{1,2}, S.~Kohle \inst{2}, M.~Hilker
\inst{2}, T.~Richtler \inst{2}, L.~Infante \inst{3}, \and H.~Quintana
\inst{3} 
} 

\offprints {M. Kissler--Patig (Bonn)}

\institute{
European Southern Observatory, Karl-Schwarzschild-Str.~2, 85748 Garching, Germany
\and 
Sternwarte der Universit\"at Bonn, Auf dem H\"ugel 71, 53121 Bonn, Germany
\and
Departamento de Astronom\'ia y Astrof\'isica, P.~Universidad Cat\'olica,
Casilla 104, Santiago 22, Chile
}

\date {}

\maketitle
\markboth{Globular cluster systems in Fornax}{}

\begin{abstract}
We studied through $V$ and $I$ photometry the properties of the globular 
cluster systems of five early--type
galaxies in the Fornax galaxy cluster: NGC 1374, NGC 1379, NGC 1387, NGC
1427, and the central giant elliptical NGC 1399.
While the four normal galaxies have between 300 and 500 globular
clusters, leading to specific frequencies of $4\pm1$ for all of them,
NGC 1399 has around $6000\pm600$ globular clusters and a specific
frequency of $12\pm3$.

The globular cluster colors are somewhat redder than those of the Milky Way 
globular clusters, however with a similar dispersion.
This indicates a wide range of metallicities in all our galaxies with a
mean metallicity somewhat higher than in the galactic halo, comparable
to the galactic bulge. In NGC 1399 the dispersion is about twice as
high, and the color distribution could be multi--modal.
No color gradient could be detected within the sensitivity of our
photometry. 

The density profiles of the globular cluster systems follow the galaxy
light in all the galaxies. In NGC 1399 the globular cluster system
is much flatter than in the other galaxies, but coincides with the large
cD envelope of the galaxy.
None of the globular cluster systems is clearly elongated, even the ones
in NGC 1374 and NGC 1427 (E1 and E3 respectively) appear rather
spherical.

From a comparison with studied globular cluster systems in spiral galaxies,
we conclude that our faint ellipticals have no more
globular clusters than spirals of the same mass, and neither differ in
their other properties, such as globular cluster colors and
morphological properties of the system.
For these galaxies there is no need to imply a different formation
scenario of the globular cluster system. The central giant 
elliptical has far more globular clusters and 
show signatures of a different history in the building up of its globular 
cluster system.


\end{abstract}

\section{Introduction}

The study of globular cluster systems around galaxies beyond the
Local Group has become routine (Harris 1991, Richtler 1995). Around 70 galaxies
and their systems have been investigated to date (Harris \& Harris 1996). 
Several systematics have been identified, but none of them
was yet fully understood with respect to the galaxy history.
The probably most controversial point is the number of globular clusters and 
the specific frequency $S$ (number of globular clusters per galaxy
luminosity in units of $M_V=-15$ mag). While the discrepancy of the $S$ values 
between spirals and ellipticals was used as an argument against a
scenario where ellipticals were built up by mergers (e.g.~ Van den Bergh 1990),
other authors (e.g.~Ashman \& Zepf 1992) suggested that the problem might be solved
by globular clusters being formed during the merging events as seen in 
several galaxies (NGC 3597: Lutz 1991; NGC 1275: Holtzman et al. ~1992; NGC
7252: Whitmore et al.~1993, Schweizer \& Seitzer 1993; He 2--10: Conti \&
Vacca 1994; NGC 4038/4039: Whitmore \& Schweizer 1995; NGC 5018: Hilker
\& Kissler--Patig 1996).
However it is still uncertain if globular cluster formation in
interactions was effective enough (e.g. Harris 1995) to explain the
apparently higher number of globular clusters in ellipticals compared to
spirals.

We investigate here
a sample of five early--type galaxies in the Fornax galaxy cluster.
Our target galaxies are NGC 1374, NGC 1379, NGC 1387,
NGC 1427, and the central giant cD galaxy NGC 1399. The globular cluster
luminosity functions have been presented in Kohle et al.~(1996,
hereafter Paper I). Earlier
photographic work already exists for these galaxies (Hanes \& Harris 
1986) and provides a basis for comparison. The globular cluster system of
NGC 1399 has been investigated by several authors (Geisler \& Forte
1990, Wagner et al.~1991, Bridges et al.~1991, Ostrov et al.~1993), to
which we will refer later on.
General data of our galaxies are summarized in Table 1.
\begin{table*}
\begin{center}
\caption{General data of our target galaxies, all members of the Fornax
galaxy cluster, taken from Tully (1988), Poulain (1988), and Poulain \& Nieto 
(1994)}
\begin{tabular}{l c c c c l l c c }
\hline
name & RA(2000) & DEC(2000) & $l$ & $b$ & type & $m_V$ & $(V-I)$ & $V_{0}$[km~s$
^{-1}$] \\
\hline
NGC 1374 & 03 35 16 & -35 13 35 & 236.36 & -54.29 & E1 & 11.20 & 1.20 &
1105 \\
NGC 1379 & 03 33 03 & -35 26 26 & 236.72 & -54.13 & E0 & 11.20 & 1.19 &
1239 \\
NGC 1387 & 03 36 57 & -35 30 23 & 236.82 & -53.95 & S0 & 10.81 & 1.30 &
1091 \\
NGC 1399 & 03 38 29 & -35 26 58 & 236.71 & -53.64 & E0 & 9.27 & 1.25 &
1294 \\
NGC 1427 & 03 42 19 & -35 23 36 & 236.60 & -52.85 & E3 & 11.04 & 1.15 &
1416 \\
\hline
\end{tabular}
\end{center}
\end{table*}
We investigated the number of globular clusters (Sect.~3),
the colors and color distributions (Sect.~4), as well as the radial and
spatial distribution of the globular clusters (Sect.~5). 
Our conclusions are presented in Sect.~6.
%

\section{Observations and reduction}

\subsection{The observations}

The data were obtained in the nights of 26--29 September,
1994 at the 100 inch telescope of the Las Campanas observatory, Chile, run by 
the Carnegie Institution. We used a Tektronix $2048\times 2048$ pixel
chip, with a pixel size of $21\mu$m or $0\farcs227$ at the
sky, corresponding to a total field of view of $7\farcm74\times
7\farcm74$. The chip was operated with a readout noise of $8.3 e^-$, and 
a gain of $3.1 e^-/$ADU.

The observations of the different galaxies are detailed in Table 2. In
addition to the long exposures, we obtained several 60--second exposures
in both filters centered on all the galaxies.
\begin{table}
\caption{The log of our observations: all exposures were approximately
centered on the galaxies, except for NGC 1399, which was covered by a 4
image mosaic. We used a Bessell $V$ and Kron-Cousins $I$ filter. The seeing 
is quoted as measured from the FWHM of stellar object on the images}
\begin{tabular}{c c r l c}
\hline
Galaxy & Filter & Obs. date & Exposure time & seeing\\
\hline
{\small NGC 1374} & $V$  & 28.9.94 & $2 \times 1200$s & 1".2 \\
         & $I$ & 29.9.94 & $1200$s +$2\times 600$s & 1".5 \\
{\small NGC 1379} & $V$  & 29.9.94 & $2\times 1200$s & 1".2 \\
         & $I$ & 26.9.94 & $3\times 1200$s & 1".4 \\
{\small NGC 1387} & $V$  & 28.9.94 & $2\times 900$s & 1".3 \\
         & $I$ & 28.9.94 & $2\times 900$s & 1".2 \\
{\small NGC 1427} & $V$  & 26.9.94 & $3\times 1200$s & 1".5 \\
         & $I$ & 26.9.94 & $3\times 1200$s & 1".3 \\
{\small NGC 1399} & $V$  & 27.+28.9.94 & $2\times 900$s / field& 1".0\\
         & $I$ & 27.+28.9.94 & $2\times 900$s / field& 1".2\\
\hline
\end{tabular}
\end{table}
The SE and SW fields around NGC 1399 will not be considered in
the following: several $I$ exposures of the SW field were corrupted,
while the SE field includes part of NGC 1404 (another member of Fornax),
which makes the allocation of globular clusters to the one or the other
galaxy confusing. 

\subsection{The reduction}

All reductions were done in IRAF. Bias frames were subtracted and sky 
flat--fields of different nights were averaged to flatten the images
better than 1\%. The different long exposures were then combined with a sigma
clipping algorithm, to remove the cosmetics from the final frames.

Object search, photometry, and the determination of the completeness
factors were done with the DAOPHOT II version in IRAF.
For all galaxies we computed an isophotal model in each filter
(using the STSDAS package {\it isophote}) that we subtracted from our final 
long exposure to obtain a flat background for the object search and photometry.

The nights of the 26, 27 and 29 were photometric and the calibration was
done via typically 15-30 standard stars from the Landolt (1992)
list, taken throughout the nights, by which our Bessell $V$ colors were
transformed to Johnson $V$.
In the middle of the night of the $28^{th}$ cirrus passed. Frames taken
at that time were calibrated via aperture photometry on the galaxy
published by Poulain (1988), and Poulain \& Nieto (1994) and cross--checked 
with overlapping
frames in the case of NGC 1399. All other calibrations were also
inter--compared and found compatible with the aperture photometry values for the
individual galaxies. Table 3 shows our final coefficients for the
calibration equations:\newline
 $ V_{inst}=V+v1+v2\cdot X_{V}+v3\cdot (V-I)$\\
 $ I_{inst}=I+i1+i2\cdot X_{I}+i3\cdot (V-I)$\\
where instrumental magnitudes are normalized to 1 second and given with
an offset of 25 mag.
\begin{table}
\caption{Calibration coefficients for our different nights. Column 5
lists the RMS of the difference between our standard magnitudes and our
calibrated ones}
\begin{tabular}{lclcc}
\hline
night  & $v1$ & $v2$ & $v3$ & RMS \\
\hline
26.9.& $1.718\pm.018$ & $0.100\pm.012$ & $-0.019\pm.007$ & 0.020\\
27.9.& $1.671\pm.009$ & $0.133$ fixed & $-0.020\pm.009$ & 0.024\\
28.9.& $1.713\pm.018$ & $0.101\pm.012$ & $-0.018\pm.007$ & 0.020\\
29.9.& $1.668\pm.009$ & $0.133$ fixed & $-0.020\pm.009$ & 0.024\\
\hline
night & $i1$ & $i2$ & $i3$ & RMS \\
\hline
26.9.& $2.077\pm.012$ & $0.038\pm.008$ & $-0.013\pm.005$ & 0.012\\
27.9.& $2.065\pm.005$ & $0.047$ fixed & $-0.017\pm.005$ & 0.011\\
28.9.& $2.069\pm.016$ & $0.041\pm.010$ & $-0.006\pm.007$ & 0.013\\
29.9.& $2.066\pm.007$ & $0.047$ fixed & $-0.015\pm.006$ & 0.015\\
\hline
\end{tabular}
\end{table}

The completeness calculations were done by standard artificial star experiments.
We added typically 10000 stars over many runs on a long exposure and
repeated the reduction steps starting with the object finding.
The completeness values are given in detail in Fig.~1 of 
Paper I. The completeness limit of 50\% is reached at
$V\simeq 23.5$ mag, $I\simeq 22.5$ mag for our four normal galaxies, 0.5
mag deeper for the NE and NW fields (fields 2 and 4 in Paper I) around
NGC 1399. 

\section{The number of globular clusters}

\subsection{Globular cluster counts}

We computed the number of globular clusters present down to the turn--over of 
the globular cluster luminosity functions (GCLF) presented in Paper I for each
galaxy. We then corrected these counts for a 
possible incompleteness in the area covered, using the density profiles
of Sect.~5.2. We assumed the GCLF to be symmetric around the turn--over and
doubled our counts to get the total number of globular clusters in the
galaxies. 

For all the galaxies, except for NGC 1399, the counts for the globular
cluster luminosity function were computed in an area extending out to
120\arcsec\ radius (about 10 kpc at the distance of Fornax). The density 
profiles in Sect.~5.2 show that the
density of globular clusters reaches the background at about this radius
in NGC 1379 and NGC 1387, but extends further in NGC 1374 and NGC 1427. 
Geometrical corrections had therefore to be applied for the non--covered areas 
in the center of the galaxies (see columns 4 and 5 in Table 4), as well
as for the region beyond 120\arcsec\ from the center of the galaxies
(column 6 in Table 4).
The latter correction was computed by subtracting the respective mean
background densities (see Sect.~5.2) from the object densities in Table 7, 
then summing the number of globular clusters in the outer rings
after correction for completness. 
For the central regions of the galaxy, we extrapolated 
the density profiles inwards taking into account that the density in a globular 
cluster system does not rise steeply towards the center (e.g.~Aguilar et 
al.~1988) and
assumed a density of $75\pm50$ globular clusters per square arcmin 
down to the turn-over in the inner, uncovered regions of all four galaxies.

For NGC 1399, we individually computed the counts on our NE and NW
fields from 1.0\arcmin\ to 9.5 \arcmin\ radius. The NE fields holds 20.1\%,
the NW field 20.8\% of this ring and includes the center of the galaxy.
For the inner region we assumed a mean density of $100\pm50$ globular
clusters per square arcmin down to the turn--over (see the density profiles
in Sect.~5.2), thus $560\pm280$ objects over the 2.8 square arcmin for
the full GCLF. 
Further we estimated the globular cluster system to extend out to about
12\arcmin\ (see also Hanes \& Harris 1986), with a mean density of $1\pm1$
globular clusters per square arcmin down to the turn--over beyond
9.5\arcmin\, thus $340\pm340$ globular clusters over the full GCLF in the 
uncovered outer region. From both the NE and NW field we then
extrapolated the total number of globular clusters around the whole
galaxy and got consistent results.

The results are summarized in Table 4
with the following sources of errors taken into account.
The number of globular clusters were computed in Paper I as the
excess objects around the galaxies, and thus affected by the
uncertainties of the background determination (square root of the
background counts) and of the turn--over value. The counts down to the 
turn--over with their errors are shown in column 2, the assumed turn--over 
stands in column 3.
\begin{table*}
\caption{Number of globular clusters around our target galaxies. 
Column 1 lists the name of the galaxy, column 2 the background corrected 
counts down to the turn--over of the GCLF together with
the errors from the background correction and the error in the assumed
turn--over. The measured turn--over value of the GCLF is shown 
in column 3, column 4 and 5 list the uncovered area towards the
center of the galaxies and the correction, column 6 lists the correction for
the area beyond 120\arcsec\ from the center, column 7 and 8 give the total
amount of globular clusters within 120\arcsec\ and around the galaxy}
\begin{tabular}{lcccclll}           
\hline
Galaxy&raw counts&turn--over&\multicolumn{2}{c}{geom.~correct.}&
correct. $>$120\arcsec & total $<$ 120\arcsec\ & {\bf total}\\
\hline
\multicolumn{6}{c}{counts from the GCLF in $V$}\\
\hline
NGC 1374 & $118\pm15\pm11$&$23.52\pm0.14$&0.072&$5.4\pm3.6$& $156\pm72$
& $248\pm38$ & {\bf $404\pm81$}\\
NGC 1379 & $114\pm15\pm24$&$23.68\pm0.28$&0.091&$6.8\pm4.6$& $51\pm19$
& $245\pm57$ &{\bf $296\pm60$}\\
NGC 1387 & $143\pm07\pm31$&$23.80\pm0.20$&0.450&$34\pm23$& $24\pm10$ 
& $365\pm110$ & {\bf $389\pm110$}\\
NGC 1427 & $166\pm19\pm21$&$23.78\pm0.21$&0.072&$5.4\pm3.6$& $189\pm66$
& $344\pm58$ & {\bf $533\pm88$}\\
NGC 1399 NE & $539\pm19\pm43$&$23.90\pm0.08$&\multicolumn{2}{c}{see text}
& --- & --- & {\bf $6270\pm640$}\\
NGC 1399 NW & $525\pm19\pm38$&$23.90\pm0.08$&\multicolumn{2}{c}{see text}
& --- & --- & {\bf $5960\pm600$}\\
\hline
\multicolumn{6}{c}{counts from the GCLF in $I$}\\
\hline
NGC 1374 & $124\pm11\pm17$&$22.60\pm0.13$&0.072&$5.4\pm3.6$& $156\pm72$
& $259\pm41$ & {\bf $415\pm83$}\\
NGC 1379 & $133\pm11\pm29$&$22.54\pm0.34$&0.091&$6.8\pm4.6$& $51\pm19$ 
& $280\pm63$ &{\bf $331\pm66$}\\
NGC 1387 & --- & --- & --- & --- & ---  & --- & ---\\
NGC 1427 & $143\pm10\pm13$&$22.31\pm0.13$&0.072&$5.4\pm3.6$& $189\pm66$
& $297\pm56$ &{\bf $486\pm87$}\\
NGC 1399 NE & $457\pm16\pm19$&$22.27\pm0.05$&\multicolumn{2}{c}{see text}
& --- & --- & {\bf $5460\pm505 $}\\
NGC 1399 NW & $537\pm16\pm24$&$22.27\pm0.05$&\multicolumn{2}{c}{see text}
& --- & --- & {\bf $6070\pm520$}\\
\hline
\end{tabular}
\end{table*}
  
These total numbers compare well with the results of the photographic work of
Hanes \& Harris (1986) who found very similar raw counts down to their limiting
magnitudes of $B_j=23.40$--$23.95$. For NGC 1399, the various older works
that determined the number of globular clusters (Wagner et al.~1991,
Bridges et al.~1991) assumed larger distances for the Fornax cluster, and 
thus extrapolated their luminosity functions to larger total numbers.
However their counts to their limiting magnitudes agree with ours.

\subsection{The specific frequency}

To derive the specific frequency (number of globular clusters per unit
galaxian luminosity in units of $M_V=-15$) we determined the distance
of the galaxies from their GCLF turn-over magnitudes. We computed the
mean of the distance moduli derived in $V$ and $I$, assuming an absolute
turn--over magnitude of $V_{TO}=-7.4\pm0.2$ and $I=-8.5\pm0.2$ (see
Paper I). The absolute magnitudes of the galaxies (see Table 1) were
then combined with the total number from the previous
subsection using the mean of the counts in $V$ and $I$. The results are 
shown in Table 5.
\begin{table}
\caption{Specific frequency for our target galaxies. Column 1 lists the
name of the galaxy, column 2 the derived distance modulus, column 3 the
absolute magnitude in $V$, column 4 the total number of globular
clusters $N_t$, column 5 the specific frequency $S_N$}
\begin{tabular}{crrlr}
\hline
Galaxy & $(m-M)$ & $M_V$ & $N_t$ & $S_N$ \\
\hline
NGC 1374 & $31.0 \pm.2$ & $-19.8\pm.2$ & $410 \pm 82$ & $4.9\pm 1.3$ \\
NGC 1379 & $31.1 \pm.2$ & $-19.9\pm.2$ & $314 \pm 63$ & $3.4\pm 0.9$  \\
NGC 1387 & $31.0 \pm.2$ & $-20.2\pm.2$ & $389 \pm 110$ & $3.2\pm 1.1$ \\
NGC 1427 & $31.0 \pm.2$ & $-20.0\pm.2$ & $510 \pm 87$ & $5.1\pm 1.3$ \\
NGC 1399 & $31.0 \pm.2$ & $-21.7\pm.2$ & $5940 \pm 570$ & $12.4\pm 3.0$ \\
\hline
\end{tabular}
\end{table}
 
NGC 1374, NGC 1379, NGC 1387, and NGC 1427 have
specific frequencies around 4.0, while NGC 1399 deviates from
that mean with $S_N=12.4\pm3.0$. Again this is in agreement with the
previous works. Hanes \& Harris (1986) gave various values of $S_N$ in
function of the distance to Fornax (affecting both their counts and the
luminosity of the galaxies) of around 4 for NGC 1374, NGC 1379, and NGC
1387. For NGC 1399 their values range from 11 to 19. Wagner et al.~(1991)
quoted a value of $15.4\pm3.2$ for NGC 1399, Bridges et al.~(1991) a
value of $16\pm4$. An interpretation will be given in the
discussion.

%
\section{The colors of the globular cluster systems}

In this section we analyze the colors of the globular clusters around
our galaxies. 
Globular clusters were selected in the same manner as for the GCLF,
i.e.~by the second moments of the intensity, as well as by the SHARP
and CHI value returned from DAOPHOT.  Most background galaxies (the
main source for contamination) have outstanding
values in at least one of these parameters, and could be removed.

\subsection{The color distributions in the normal galaxies}

We again considered only objects closer than 120\arcsec\  to the 
center of the respective galaxy in the case of our four normal galaxies.
\begin{figure}
\psfig{figure=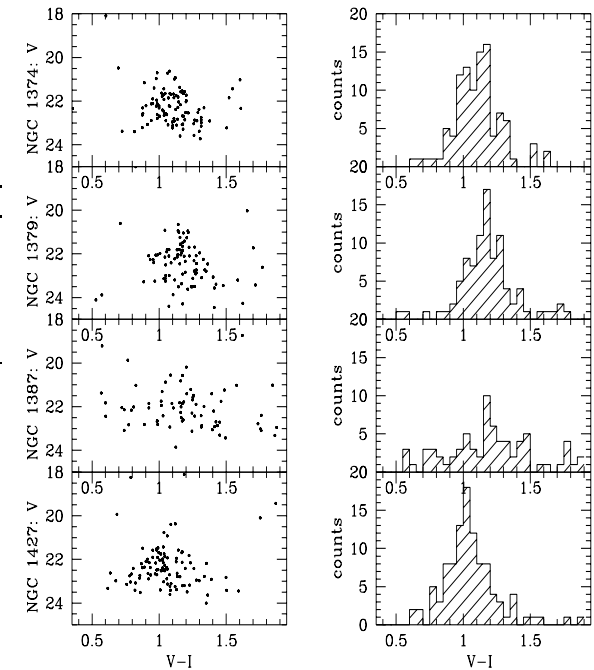,height=9cm,width=8cm
,bbllx=8mm,bblly=57mm,bburx=205mm,bbury=245mm}
\caption{
The color distribution of all globular clusters 
around our galaxies. On the left our color magnitude diagrams, on the
right the corresponding histograms with all globular clusters included 
}
\end {figure}
\begin{figure}
\psfig{figure=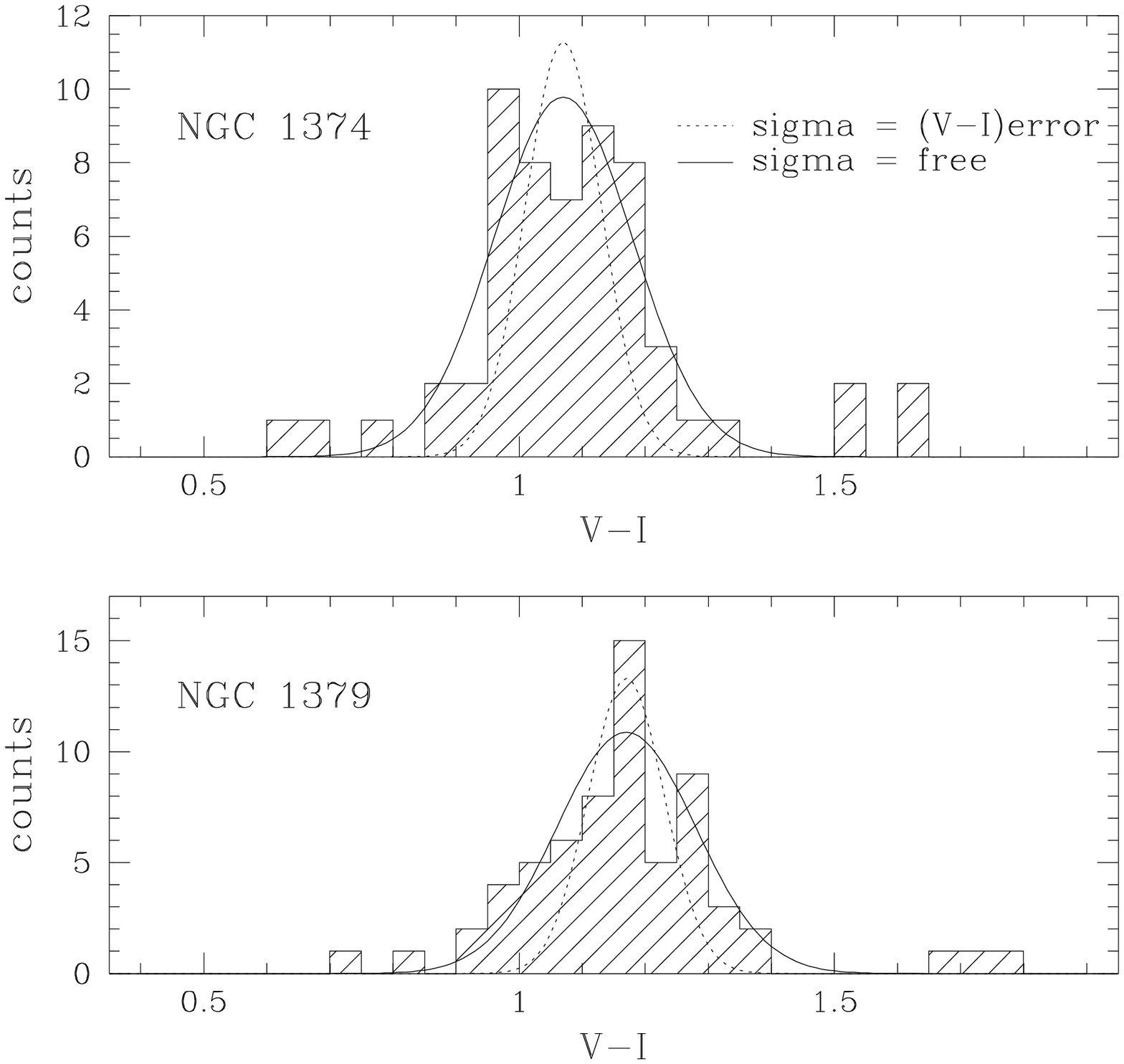,height=9cm,width=8cm
,bbllx=8mm,bblly=57mm,bburx=205mm,bbury=245mm}
\psfig{figure=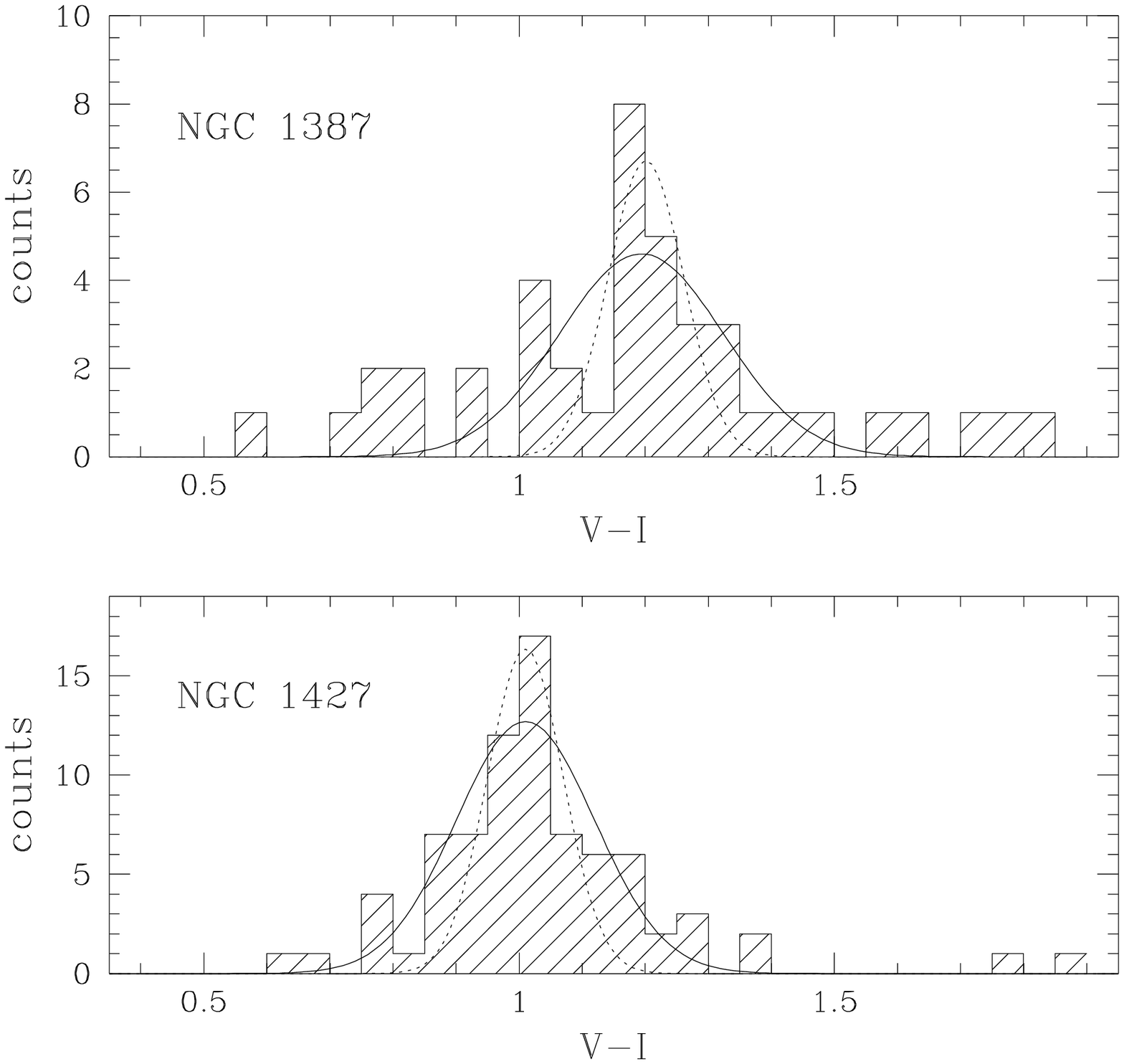,height=9cm,width=8cm
,bbllx=8mm,bblly=57mm,bburx=205mm,bbury=245mm}
\caption{Histogram of the globular cluster color distribution in $(V-I)$ 
around NGC 1374, NGC 1379, NGC 1387, and NGC 1427. The solid line is the
gaussian fit to the distribution, the dotted one shows the broadening
from the errors in $(V-I)$ alone
%
}
\end {figure}
The color distribution of all the globular clusters around NGC 1374, NGC 1379,
NGC 1387, and NGC 1427 are shown in Fig.~1. The same distribution for
globular clusters that have errors in $V-I$ less than 0.1 mag (typically
occurring between $V=22.5$ mag and $V=23.0$ mag) are shown in Fig.~2,
in which we over--plotted two curves: \\
$\bullet$ the best fitting gaussian (solid line), which returned
widths of $\sigma = 0.11$ to $0.13$ mag.\\
$\bullet$ the expected distribution (dotted line) if all globular clusters
would have the same color, and the broadening would be due to the
errors of the photometry alone.\\
The narrow dispersions of our free gaussian fits show that the {\it intrinsic}
dispersion of the distributions must be less than
0.1 mag. We performed the KMM test proposed by Ashman et al.~(1994), 
as well as fits with multiple gaussians, in order to detect
multi--modality in the distributions, but in no case the hypothesis of an
unimodal distribution could be rejected.
 
While the width of the distribution is almost identical for these four
galaxies, the median color slightly differs. We assumed $E(B-V)=0.0$ towards
Fornax (Burstein \& Heiles 1982), and derived median $(V-I)$ colors of the 
globular clusters shown in Table 6, together with a median metallicity.
 
Deriving metallicities from broadband colors is complicated by second
parameter effects: age and metallicity can hardly be disentangled. 
For similar ages and metallicities, the $V-I$ color of
galactic globular clusters spread over about 0.2 mag. 

From the relatively red ($V-I > 0.8$) mean colors, the quite narrow
color distributions, the fact that we see no peculiarities in the
luminosity functions of the globular clusters, and that there is no sign for any
recent mergers, we are encouraged to assume that the globular clusters in these
four galaxies are older then 10 Gyr.  With this assumption on age
we can roughly convert our colors to metallicities
using the empirical color--to--metallicity relation found for the Milky Way 
globular clusters. 

We updated the color--metallicity relation given in Couture et al.~(1990), using
the Mc Master catalogue (Harris 1996) of Milky Way globular clusters.
Figure 3 shows the metallicity of all the galactic globular clusters with a 
reddening $E(B-V)$ less than 0.4 (60 candidates) plotted against their
$(V-I)_0$ color, that we corrected for extinction assuming $E(V-I) = 1.38
\times E(B-V)$ (Taylor 1986).
\begin{figure}
\psfig{figure=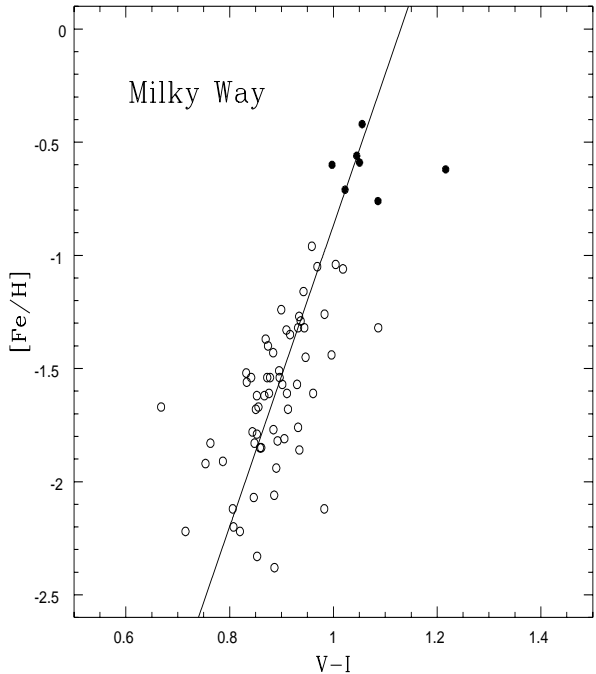,height=9cm,width=8cm
,bbllx=8mm,bblly=57mm,bburx=205mm,bbury=245mm}
\caption{Metallicity versus $(V-I)$ color for the globular clusters in
the Milky Way with $E(B-V) < 0.4$. Dots are bulge clusters, circles are
halo clusters, the line shows the best linear fit
}
\end {figure}
The best linear fit gives the relation:\\
$(V-I)_0 = 0.15 (\pm 0.02) [$Fe/H$] + 1.13 (\pm 0.03)$\\
This is valid, strictly speaking, only in the range $0.7 < (V-I) < 1.1$.
\begin{table}
\begin{center}
\caption{Median $V-I$ colors of the globular clusters around NGC 1374,
NGC 1379, NGC 1387, and NGC 1379; as well as derived median
metallicities}
\begin{tabular}{ccr}
\hline
Galaxy & median $(V-I)$  & [Fe/H]\\
\hline
NGC 1374 & $1.10 \pm 0.03$  & $-0.2\pm0.3$\\
NGC 1379 & $1.17 \pm 0.03$ & $0.3\pm0.3$\\
NGC 1387 & $1.20 \pm 0.06$ & $0.5\pm0.4$\\
NGC 1427 & $1.03 \pm 0.03$ & $-0.7\pm0.3$\\
\hline
\end {tabular}
\end{center}
\end{table}
A direct comparison to the colors of the globular clusters in the Milky
Way is also of interest. The color distribution of all the globular
clusters in the Milky Way is shown in Fig.~4, the contribution of halo 
and bulge globular clusters are over--plotted (dashed and black).
\begin{figure}
\psfig{figure=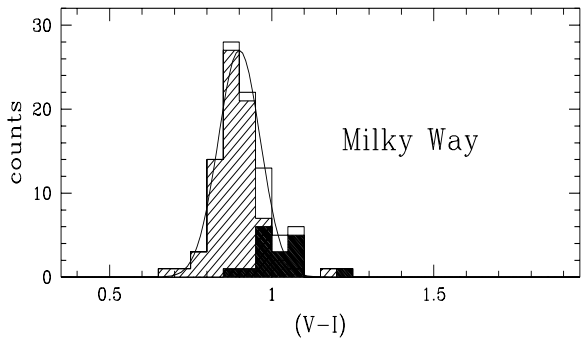,height=4cm,width=8cm
,bbllx=8mm,bblly=157mm,bburx=205mm,bbury=240mm}
\caption{Histogram over the $(V-I)$ color of the globular clusters in
the Milky Way. The dashed region marks the contribution of the halo
clusters, the black ones that of the bulge
}
\end {figure}
A gaussian fit to the histogram leads to a mean value of $(V-I) = 0.90
\pm 0.01$ and a dispersion of $\sigma = 0.065 \pm 0.003$.
This is comparable to the {\it intrinsic} dispersion of the
color distributions of our galaxies, and corresponds to a range of metallicity 
from $[Fe/H] = -2.4$ to $0.2$ dex.

Thus, from the colors alone, accurate metallicities cannot be derived,
but we can conclude that:
\begin{itemize}
\item All our early-type galaxies have median colors of their globular
clusters redder than that of the Milky Way. Assuming old globular cluster
populations, this would mean average metallicities of the globular
clusters ranging from slightly richer than the Milky Way halo (for NGC 1427)
to about solar (for NGC 1379 and NGC 1387).
\item There is no clear evidence for multiple populations, the range of
metallicities for the globular clusters in each galaxy could span 2
dex in [Fe/H] as in the Milky Way, but is concentrated around a median
value.
\item The median color for globular clusters in all the galaxies is bluer
by about 0.1 mag in $(V-I)$ than the integrated galaxy light, as already 
noticed in all galaxies observed up to date,
but all our galaxies possess also globular clusters as red as the galaxy
itself.
\end{itemize}
 
\subsection{The globular cluster colors in NGC 1399}
 
For NGC 1399 we considered all the globular clusters in the NE and NW
fields obeying our selection criteria.
Figure 5 shows the color distribution of the
globular clusters with errors in $V-I$ less than 0.1 mag. 
Here the best gaussian fit returns a width of $\sigma = 0.22 $,
three times as large as expected from the errors only. 
For the sample composed by all globular clusters in the NE and NW field
the KMM test rejects the unimodal hypothesis (with a confidence over 95\%)
if we impose dispersions of the color distributions comparable to the 
dispersions observed in our normal galaxies ($\sigma\simeq 0.12$), and
favors two populations centered on $V-I=0.99$ and $V-I=1.18$
(corresponding to [Fe/H]$=-0.9$ and $0.3$ dex according to our relation
in the previous section). 
We note that the globular clusters do not span a much wider range of
colors than in the other galaxies, but rather populate all colors more
homogeneously, i.e.~they are no globular clusters with
peculiar colors in NGC 1399 compared to the other galaxies.

From Washington photometry Ostrov et
al.~(1993) also derived multiple peaks in the color distribution. They
find two groups, at [Fe/H]$=-1.5$ and $-0.9$, as well as a possible third
group near $-0.2$ dex. The shift in the color to metallicity conversion
between our results demonstrates how difficult metallicity estimates are
from broad band colors alone. The results from Washington photometry are
probably more reliable than from the less metal sensitive $V,I$ colors.
The main result here is the confirmation of at least two groups of globular
clusters in the color distribution. 
\begin{figure}
\psfig{figure=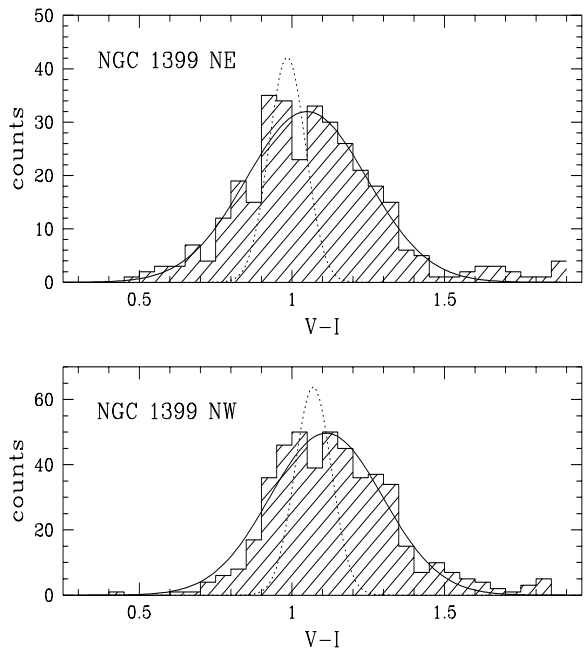,height=9cm,width=8cm
,bbllx=8mm,bblly=57mm,bburx=205mm,bbury=245mm}
\caption{
Color distributions for the globular clusters in the NE (upper panel)
and NW (lower panel) fields around NGC 1399. The solid line is a free
gaussian fit and returns a dispersion of 0.22, while the dotted line
shows the broadening expected from our errors in photometry alone
}
\end {figure}

This would suggest that two or more distinct globular cluster enrichment 
or formation epochs/mechanisms happened in NGC 1399. 

\subsection{A composite color distribution}

As an experiment we constructed an artificial color
distribution with all globular clusters found around NGC 1374, NGC 1379,
NGC 1387, and NGC 1427 with errors in $V-I < 0.1$ mag. This is equivalent
to the sum of the four histograms shown in Sect.~4.1.

The resulting composite color distribution is shown in Fig.~6. We
performed the same test on it as for the other distributions: a
single gauss fit returns a width of $\sigma=0.31$; the KMM test favors a
double gauss fit with 99\% confidence if we impose individual width of
$\sigma=0.12$.  
\begin{figure}
\psfig{figure=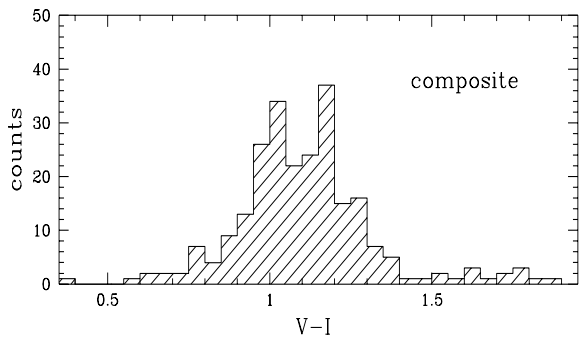,height=4cm,width=8cm
,bbllx=8mm,bblly=157mm,bburx=205mm,bbury=240mm}
\caption{Composite histogram of the $(V-I)$ colors of all the globular 
clusters with a $V-I$ error less than 0.1 mag in NGC 1374, NGC 1379, NGC 1387,
and NGC 1427. The distribution seems to be bi--modal but is based on
four systems
}
\end {figure}
This color distribution would probably be classified as {\bf bi--}modal, in
case of real data, while it is composed of {\bf four} globular cluster
systems. We take it as a word of caution, that broad and multi--modal
color distribution might hide a much more complex history than a single
merger event between two galaxies, as Ashman \& Zepf (1992) propose in a
first approximation. 

\subsection{Color gradients}

We present the radial color distributions of the globular clusters in
Fig.~7. The globular clusters plotted are the same as used for the
color distributions in the sections above, i.e.~objects that match our
criteria for point like sources have an error in $V-I$ of less
than 0.1 mag, and are closer than 120\arcsec\  (425\arcsec\  in the case of
NGC 1399) to the center of the parent galaxy.
\begin{figure}
\psfig{figure=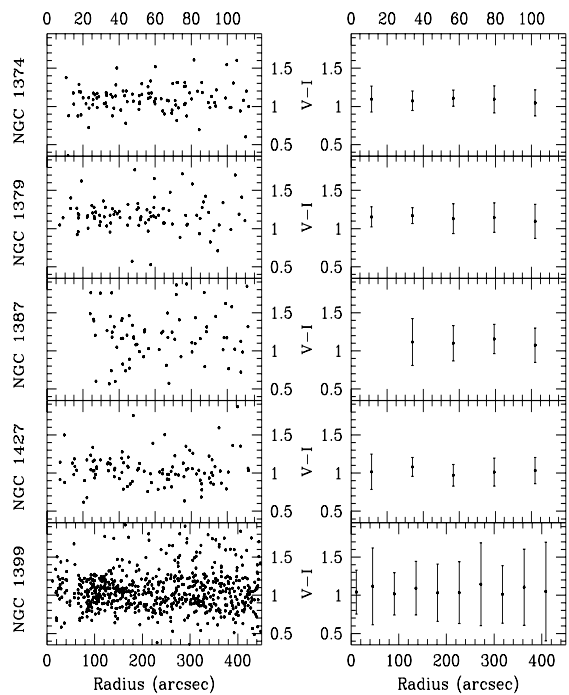,height=9cm,width=8cm
,bbllx=8mm,bblly=57mm,bburx=205mm,bbury=245mm}
\caption{
Color gradients for our five target galaxies. Plots extend to 120\arcsec\
for NGC 1374, NGC 1379, NGC 1387, and NGC 1427; to 425\arcsec\ for NGC
1399. Errorbars show the dispersion around the median
}
\end {figure}
Color gradients in globular cluster systems is a topic of much
debate. Where they are found, such gradients are small, typically 
of the order of 0.1 to 0.3 mag in common broad band
indices over hundreds of arcseconds radius (e.g.~in M87, Lee \& Geisler 1993,
in M49 and NGC 4649, Couture et al.~1991, in NGC 3923 Zepf et al.~1995,
or in NGC 3311 Secker et al.~1995). 
For most galaxies where such a gradient is found, another study exists quoting 
a non--detection.

The most extensive data to date are probably from Geisler et al.~(1996), who 
recently re--examined the globular cluster system of
M49 with deep Washington photometry, and clearly showed a color gradient. 
However, the gradient is rather due to a different mixture of two populations in
the inner and outer parts than to a steady increase of metallicity to
the center as interpreted by authors in the past.

The gradient in M49 is of the order of 0.4 $\Delta $[Fe/H]/$\Delta log R$.
This would translate into 0.06 $\Delta(V-I)/\Delta log R$ according to
our relation of Sect.~4.1, or $\Delta(V-I)=0.12$ to 0.16 mag over the
range studied in our cases. We are therefore clearly not sensitive
enough to detect such gradients in our data, and can only exclude
gradients as large a 0.15 $\Delta(V-I)/\Delta log R$ (or 1.0 $\Delta
[Fe/H]/\Delta log R$) for our galaxies. One would need very good photometry
a couple of magnitudes deeper in a sensitive system (e.g.~Washington or 
$B$ and $I$
Johnson--Cousins, Geisler et al.~1996) to find gradients, if present, in
the globular cluster systems studied here.

Bridges et al.~(1991) found a decrease of 0.2 mag in $B-V$ from 1\arcmin\
to 3\arcmin\  of the center in NGC 1399. Over this range, the increase
might also exists in our data, but is smaller than the scatter and
can in no case be extrapolated further out.   

An interesting characteristic of the globular clusters in NGC
1399 might be noticed: the large dispersion of colors derived in 
Sect.~4.2 exists at all radii, as in the case of M49 (Geisler et al.~1996).

%
\section{Spatial distributions}

\subsection{The angular distributions}

We looked for any anisotropic distribution of globular clusters around 
NGC 1374, NGC 1379, NGC 1387, and NGC 1427. No deep image centered on NGC
1399 was available.
We computed the counts in the same rings as for the GCLFs
(i.e.~excluding the centers, and out to 120\arcsec\ ). We devided the ring
in $16\times22.5$ degrees segments around NGC
1374, NGC 1379, NGC 1387, and NGC 1427, and plotted the distributions
modulo $\pi$, (i.e.~rotating the western side by 180 degrees around the
center to increase a possible excess along a given axis) in Fig.~8.
For NGC 1374 and NGC 1427 (E1 and E3 galaxies respectively), we
indicated the position angle of the galaxies with dotted lines. The
amount of background contamination in a segment is shown as a solid line. 

All the distributions are compatible with the globular clusters being
spherically distributed around the galaxy. In NGC 1374 a $2 \sigma$ excess of
objects along the major axis of the galaxy is present. For NGC 1427
it can be excluded that the globular cluster system is as elliptical as
the galaxy. For NGC 1379 (E0), the distribution with an 
ellipticity of $0.2\pm0.1$ and a position angle of $70\pm10$ degrees fits
the data equally well as a spherical distribution.
\begin{figure}
\psfig{figure=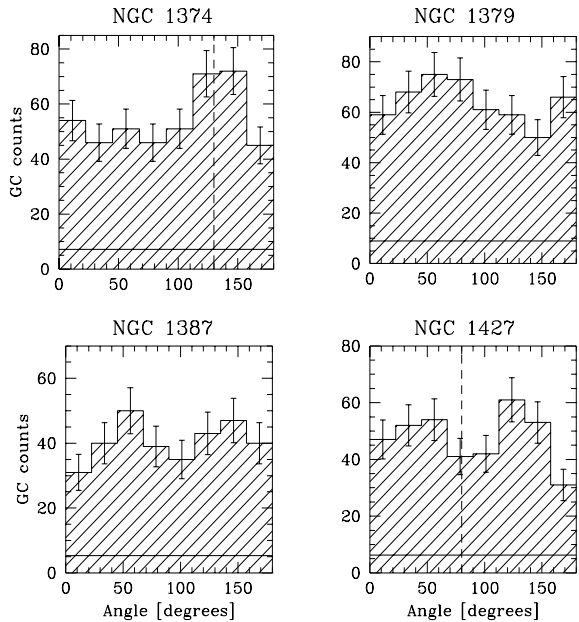,height=8cm,width=8cm
,bbllx=8mm,bblly=57mm,bburx=205mm,bbury=245mm}
\caption{
The angular distributions of globular clusters around NGC 1374, NGC
1379, NGC 1387, and NGC 1427. The globular clusters were counted in 22.5
degrees wide segments around the galaxy and taken modulo $\pi$  
}
\end{figure}

\subsection{The radial distributions}
\subsubsection{The ``normal'' galaxies}
For NGC 1374, NGC 1379, NGC 1387, and NGC 1427 we computed the surface density
profile for all objects found around the galaxy down to $V=24.0$ mag
without any correction for completeness. 
Table 7 shows the densities computed in increasing
elliptical rings 22.7\arcsec\ (100 pix) wide. The density profiles are
plotted in Fig.~9, the upper panel showing the uncorrected 
distribution, the lower panel showing the distribution corrected for
background contamination (see Sect.~5.3) together with the arbitrarily shifted 
galaxy light profile (squares).  
The excess in NGC 1374 at about 150\arcsec\ corresponds to the distance
of NGC 1375, the smaller galaxy close in projection to NGC 1374, which
might contribute a few globular clusters. 
\begin{table}
\caption{Density profiles for the normal galaxies. Column 1 shows the
semi--major axis at the center of the ring in arcsecs, columns 2--5 the 
density of objects per square arcmin for our four galaxies}
\begin{tabular}{l r r r r}
\hline
 $a$ & NGC 1374 & NGC 1379 & NGC 1387 & NGC 1427 \\
\hline
57\arcsec\ & $45.4\pm 6.1$ & $37.9\pm 5.3$ & $33.4\pm 5.0$ & $59.4\pm 8.0$ \\
80\arcsec\ & $19.8\pm 3.1$ & $21.4\pm 3.1$ & $15.2\pm 2.6$ & $21.6\pm 3.7$ \\
102\arcsec\ & $12.4\pm 2.1$ & $12.1\pm 2.0$ & $10.2\pm 1.8$ & $17.3\pm 2.8$ \\
125\arcsec\ & $7.4\pm 1.4$ & $6.9\pm 1.3$ & $5.2\pm 1.1$ & $13.8\pm 2.2$ \\
147\arcsec\ & $10.6\pm 1.5$ & $5.7\pm 1.1$ & $5.3\pm 1.0$ & $7.8\pm 1.5$ \\
170\arcsec\ & $4.0\pm 0.9$ & $7.0\pm 1.2$ & $3.6\pm 0.8$ & $10.1\pm 1.7$ \\
193\arcsec\ & $6.4\pm 1.0$ & & $4.1\pm 0.8$ & $11.4\pm 1.9$ \\
216\arcsec\ & $5.4\pm 0.9$ & & $4.0\pm 0.8$ & $9.4\pm 1.7$ \\
238\arcsec\ & $3.5\pm 0.7$ & & $6.2\pm 1.0$ & $6.2\pm 1.4$ \\
261\arcsec\ & $4.0\pm 1.1$ & & $4.4\pm 1.1$ & $6.6\pm 1.4$ \\
\hline
\end{tabular}
\end{table}
\begin{figure*}
\psfig{figure=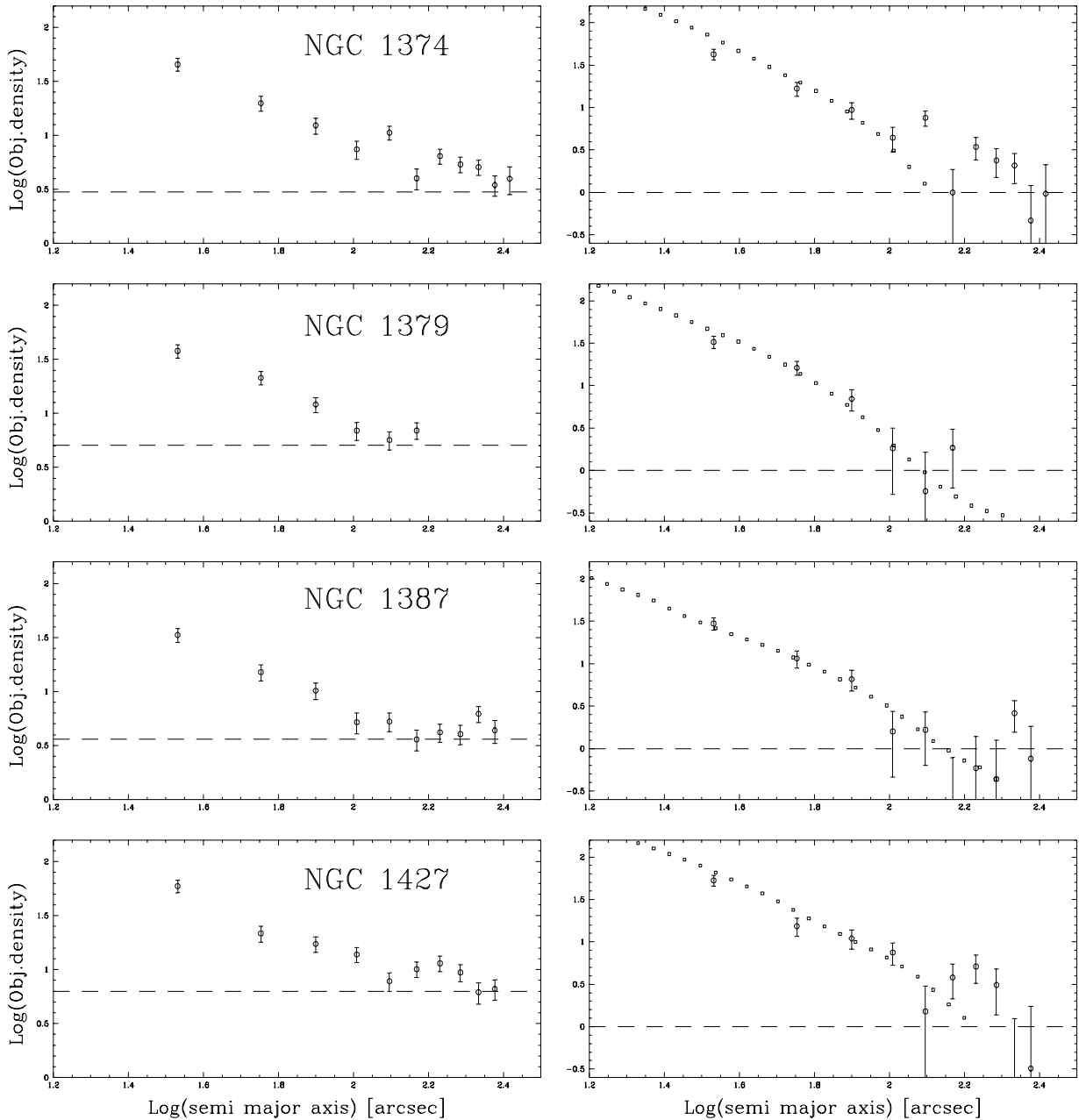,height=18cm,width=18cm
,bbllx=8mm,bblly=57mm,bburx=205mm,bbury=245mm}
\caption{
The radial distribution of objects in NGC 1374, NGC 1379, NGC 1387, and
NGC 1427. The left and right panels respectively show
the uncorrected density profile and the density profile corrected for
background contamination together with the arbitrarily shifted galaxy light 
profile (squares). Surface densities are given in number per square
arcminute
}
\end{figure*}
%
%
%
\subsubsection{NGC 1399}

For NGC 1399 we computed the surface density profile in the NE and NW fields
individually. We counted all objects found in the fields down to $V=24.0$ mag,
in rings around the center of the galaxy, 22.7\arcsec\ (100 pix) wide,
and corrected the counts for geometrical incompleteness, i.e.~we computed
the fraction of the ring seen in the field, and scaled the counts up to
the total area.
No correction for completeness of the counts was necessary, since for NGC 1399
the counts were almost complete down to the considered magnitude. The
background contamination was determined with a background field located
about 40 arcmin east. The background field needed a small ($< 5\%$)
correction for completeness to be adjusted to the fields around NGC 1399.
The result gave 361 objects on 59.0 square arcmin in the background
field down to $V=24$ mag, or a background density of $6.1\pm0.3$ objects per 
square arcmin as background value for the counts in both fields
around NGC 1399.
Table 8 shows the densities computed in increasing elliptical rings 
22.7\arcsec\ (100 pix) wide, plotted in Fig.~10.

\begin{table}
\begin{center}
\caption{The density profile of globular clusters around NGC 1399. Column
one list the mean ring radii, column 2 and 3 show the density of objects
found in the NE and NW field}
\begin{tabular}{l r r}
\hline
radius & NGC 1399 NE & NGC 1399 NW \\
\hline
  57\arcsec  &    $89.7\pm18.7$  &    $83.4\pm8.6$ \\
  80\arcsec  &    $49.7\pm10.1$  &    $34.5\pm5.8$ \\
  102\arcsec  &    $43.8\pm7.9$  &    $40.2\pm5.7$ \\
  125\arcsec  &    $38.5\pm6.4$  &    $43.1\pm5.4$ \\
  147\arcsec  &    $32.0\pm5.3$  &    $26.8\pm4.0$ \\
  170\arcsec  &    $24.5\pm4.2$  &    $25.2\pm3.6$ \\
  193\arcsec  &    $20.6\pm3.6$  &     $19.7\pm3.0$ \\
  216\arcsec  &    $14.2\pm2.8$  &    $14.0\pm2.4$ \\
  238\arcsec  &    $22.3\pm3.3$  &     $12.4\pm2.2$ \\
  261\arcsec  &    $15.3\pm2.6$  &    $11.4\pm2.0$ \\
  284\arcsec  &    $16.0\pm2.5$  &    $13.5\pm2.1$ \\
  307\arcsec  &    $12.1\pm2.1$  &    $10.2\pm1.8$ \\
  329\arcsec  &    $12.2\pm2.0$  &    $11.5\pm1.8$ \\
  352\arcsec  &    $10.7\pm1.8$  &    $11.3\pm1.8$ \\
  375\arcsec  &    $11.2\pm1.8$  &     $8.1\pm1.4$ \\
  397\arcsec  &    $13.2\pm1.9$  &    $8.2\pm1.4$ \\
  420\arcsec  &    $9.3\pm1.6$  &    $8.0\pm1.4$ \\
  443\arcsec  &    $8.8\pm1.5$  &    $8.8\pm1.4$ \\
  465\arcsec  &    $9.7\pm1.5$  &    $8.4\pm1.7$ \\
  488\arcsec  &    $6.7\pm1.3$   &   $6.0\pm1.7$ \\
  511\arcsec  &    $6.5\pm1.5$  &    $9.5\pm2.4$ \\
  534\arcsec  &    $5.9\pm1.6$  &    $6.9\pm2.3$ \\
  556\arcsec  &    $4.8\pm1.6$  &    $8.6\pm3.0$ \\
\hline
\end{tabular}
\end{center}
\end{table}
\begin{figure}
\psfig{figure=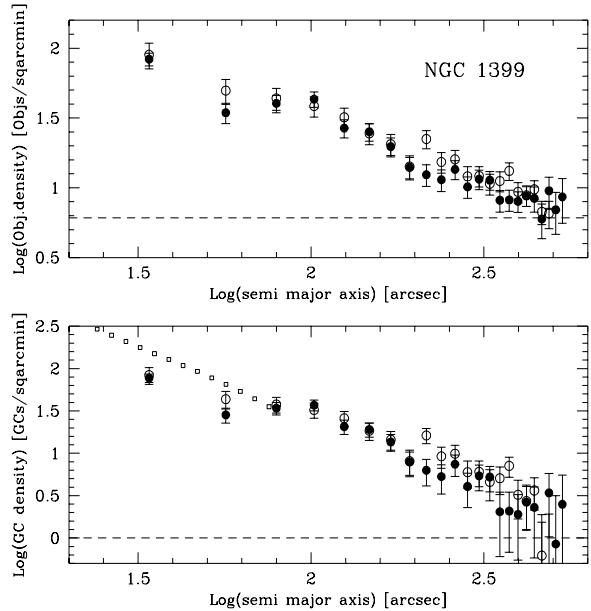,height=8cm,width=8cm
,bbllx=8mm,bblly=57mm,bburx=205mm,bbury=245mm}
\caption{
The radial distribution of objects in NGC 1399. The symbols are the same
as in Fig.~9
}
\end {figure}

\subsection{Globular cluster system vs.~galaxy profiles}

For NGC 1399, we fitted a power--law of the kind $\rho \sim r^{-x}$, 
where $\rho$ stands
for the surface density of the globular clusters or the surface
intensity of the galaxy light respectively,
and $r$ for the semi--major axis, of both the globular cluster density
profile and the light of the galaxy (taken from our isophotal models, see
Sect.~2.2, all in good agreement with similar data from Goudfrooij et al.~1994).
For the four other galaxies we solved for the slope and the background
density $\rho _{bkg}$ simultaneously by fitting a function of the kind $\rho = a\cdot
r^{-x} + \rho _{bkg}$ as proposed by Harris (1986), when the data do
not reach reliably the background level. 
The resulting coefficients are shown in Table 9. The results agree with
the data from Hanes \& Harris (1986) who found all profiles 
compatible with a slope of $-2$, and for NGC 1399 with the profile from
Wagner et al.~(1991), who found a slope of $-1.54\pm0.15$ for the
globular cluster density profile, and $-1.67\pm0.12$ for the galaxy
light.
\begin{table}
\begin{center}
\caption{Slope coefficients for the globular cluster density (column 2)
and galaxy light (column 3) of our target galaxies (column1), as well as
the fitted background (column 4)}
\begin{tabular}{l c c l}
\hline
Galaxy name & density slope & gal. light slope & bkg density\\
\hline
NGC 1374 & $1.8\pm0.3$ & $2.0\pm0.1$ & $3.0\pm2.0$\\
NGC 1379 & $2.1\pm0.6$ & $2.2\pm0.1$ & $5.1\pm2.1$\\
NGC 1387 & $2.2\pm0.3$ & $2.2\pm0.1$ & $3.6\pm0.8$\\
NGC 1427 & $2.0\pm0.3$ & $1.8\pm0.1$ & $6.3\pm1.7$\\
NGC 1399 NE & $1.55\pm0.25$ & $1.75\pm0.10$ & $(6.1\pm0.3)$\\
NGC 1399 NW & $1.75\pm0.30$ & $1.75\pm0.10$ & $(6.1\pm0.3)$\\
\hline
\end{tabular}
\end{center}
\end{table}
All density profiles of the globular clusters follow the galaxy light within
the uncertainties. But in NGC 1399 the clearly flatter globular cluster
density profile only agrees with the light profile due to the large cD
envelope of the galaxy (e.g.~Schombert 1986).

For the four normal galaxies, the values found are very similar to
results of previous studies of globular cluster systems around normal 
early-type galaxies (e.g.~Kissler--Patig et al.~1996 and references
therein).

%
\section{Discussion}

The previous sections demonstrate that while the globular cluster
systems of our faint early-type galaxies have very similar properties, 
the globular clusters in NGC 1399, the central giant elliptical cD galaxy, 
are much more numerous and have a different color distribution as well as a 
flatter density profile. 

While it is true for NGC 1399 that globular clusters appear in a
much larger number than in spirals, it is not for our fainter galaxies.
Harris \& Harris (1996) compiled all the globular cluster systems
investigated to date. If we select from their list all the S0, Sa, and
Sb galaxies (excluding the two outstanding galaxies with $M_V < -22$), we
get for the ten remaining galaxies an average number of globular clusters of 
$345\pm185$ per galaxy, for an average luminosity of $M_V=-21.1$.
The three ellipticals and the S0 galaxy that we investigated here have a mean
of $406\pm81$ globular clusters and therefore do not
have more globular clusters in absolute numbers than do these spirals.

Comparing the specific frequencies of spirals to that of ellipticals is
very difficult, if it makes sense at all. First because ideally it
should relate the number of globular clusters to the mass of the galaxy
by assuming a constant $M/L$ ratio, which is a reasonable assumption when
comparing ellipticals among each other but not when comparing spirals
with ellipticals. Second because even when reducing this discrepancy by 
normalizing the number of globular clusters to the spheroid luminosity,
it is unclear which fraction of the globular clusters in spirals are
associated with the halo and the bulge, while elliptical galaxies are
most probably bulge dominated. Thus it is not clear if we compare
comparable values. However, as a comparison, the sample of spirals
mentioned above has an average $S$ of $1.3\pm0.8$, and would $S$ be
computed for the spheroid luminosities, it would increase by about 1 
(e.g.~Harris 1991). The value for spirals does therefore not deviate
that much from the values derived in Sect.~3.2. 

Similarity seems to exist further in the color distribution of the globular 
clusters in our faint galaxies and in spirals. They are slightly redder 
(i.e.~probably more metal-rich), but show a similar dispersion around the 
median to that in the Milky Way, and cover a similar range of colors. Here
again dominating bulge clusters, in contrast to the halo
dominated Milky Way system, could possibly explain the small color differences. 

Finally we conclude that for our faint
elliptical galaxies there is no strong need to a different globular cluster 
formation or evolution scenario, as well as no need for any increase of the 
number of globular clusters during a hypothetic merger event.

On the contrary, for NGC 1399 these conclusions are not true. NGC 1399 has
far more globular clusters, and a much higher specific frequency than
the spiral galaxies. The surface density profile is much flatter and the 
globular clusters cover rather homogeneously the
full range of colors, and show signs of several populations. As
pointed out by several authors before, the formation of the globular
cluster system in NGC 1399 must have undergone a different history,
similar to other central giant ellipticals (e.g.~Harris 1991). Note that
the globular cluster system of NGC 1399
confirms all the predictions that Ashman \& Zepf (1992) made for a
globular cluster system that experienced a merger: it has a broad
(multi--modal?) color distribution, a flat surface density profile, and an
increased number of globular clusters. 
However, NGC 1399 is one of the galaxies with an outstanding specific
frequency, even for a possible enrichment by a merger. 
While the formation of a large number of globular clusters in cooling
flows seems to be ruled out (Bridges et al.~1996), it was speculated
that NGC 1399's position in the center of the Fornax cluster favored the huge
number of globular clusters also observed in other
galaxies lying at the center of galaxy clusters (e.g.~Harris 1991).
One possibility would be the increased number of merger events at early
times, since we showed that the multi--modal color distribution does not
exclude {\it several} components to have formed the globular cluster
system of NGC 1399.
Another hypothesis could be that the large number of globular clusters
is related to the large number of dwarf galaxies 
whose density Hilker et al.~(1995) reported to increase significantly towards
the center of the Fornax cluster. One could speculate that accreted while still
gaseous, the dwarf galaxies formed with high efficiency globular
clusters in the dense environment of NGC 1399. The high specific
frequency would then be a consequence of a Searle \& Zinn (1978)
scenario combined with the dense environment of NGC 1399. However, no
more than speculations could be made to date to explain the high
specific frequencies of central galaxies.

Finally we note the constancy of the specific frequencies in all our
faint galaxies. The mean for our faint early--type galaxies in Fornax is 4.2 
with a dispersion of 1.0. We can add NGC 1404, another probable member
of Fornax from a study of Richtler et al.~(1992, however note their
possible argument against a membership of the galaxy to the cluster), 
and assume a similar distance modulus of $31.0\pm0.2$. We then get
a absolute magnitude of $-21.0\pm0.2$, and a specific frequency of
$3.5\pm0.8$. The effectiveness in globular cluster formation within the normal
galaxies of the Fornax galaxy cluster must have been very similar and
might hint to similar formation histories of the galaxies in the
cluster. 

%
\acknowledgements
We wish to thank the staff of the Las Campanas observatory for the
friendly atmosphere and their valuable help during the observing run.
Thanks also to Bill Harris for providing a electronic copy of his
globular cluster system compilation, and later for his comments as
referee that helped to improve the paper.
MKP aknowledges a studentfellowship at
the European Southern Observatory, SK and MH were supported by the DFG
project Ri 418/5-1, LI would like to acknowledge
support from {\it Proyecto} FONDECYT \# 1960414.
This research made use of the NASA/IPAC extragalactic database (NED) which is 
operated by the Jet Propulsion Laboratory, Caltech, under contract with the
National Aeronautics and Space Administration.

\enddocument